\documentclass[mathleft
]{an}
\usepackage{graphicx}
\usepackage{times}
\begin{document}

\Pagespan{999}{999}
\Yearpublication{2015}%
\Yearsubmission{2015}%
\Month{999}%
\Volume{999}%
\Issue{999}%

\title{Fast outflows in broad absorption line quasars and their connection with CSS/GPS sources}

   \author{G.Bruni \inst{1,2}\fnmsep\thanks{Corresponding author. \email{bruni@mpifr-bonn.mpg.de}\newline}
 	 \and K.-H. Mack\inst{2}
	 \and F.M. Montenegro-Montes\inst{3}
	 \and M. Brienza \inst{4,5}
       \and J.I. Gonz\'alez-Serrano \inst{6}
}

\titlerunning{Fast outflows in BAL QSOs and their connection with CSS/GPS sources}
\authorrunning{G. Bruni et al.}

   \institute{Max Planck Institute for Radio Astronomy, Auf dem H\"ugel 69, D-53121 Bonn, Germany
	 \and INAF - Istituto di Radioastronomia, Via P. Gobetti 101, I-40129 Bologna, Italy
	 \and European Southern Observatory, Alonso de C\'ordova 3107, Vitacura, Casilla 19001, Santiago de Chile, Chile
	 \and Netherlands Institute for Radio Astronomy, Postbus 2, 7990 AA, Dwingeloo, The Netherlands
       \and Kapteyn Astronomical Institute, Rijksuniversiteit Groningen, Landleven 12, 9747 AD Groningen, The Netherlands
       \and Instituto de F\'isica de Cantabria (CSIC-Universidad de Cantabria), Avda. de los Castros s/n, E-39005 Santander, Spain
		}

\received{30 June 2015}
\accepted{07 Sep 2015}
\publonline{later}



\abstract{Broad absorption line quasars are among the objects presenting the fastest outflows. The launching mechanism itself is not completely understood. Models in which they could be launched from the accretion disk, and then curved and accelerated by the effect of the radiation pressure, have been presented. We conducted an extensive observational campaign, from radio to optical band, to collect information about their nature and test the models present in the literature, the main dichotomy being between a young scenario and an orientation one. We found a variety of possible orientations, morphologies, and radio ages, not converging to a particular explanation for the BAL phenomenon. \\
From our latest observations in the m- and mm-band, we obtained an indication of a lower dust abundance with respect to normal quasars, thus suggesting a possible feedback process on the host galaxy. Also, in the low-frequency regime we confirmed the presence of CSS components, sometime in conjunction with a GPS one already detected at higher frequencies. Following this, about 70\% of our sample turns out to be in a GPS or CSS+GPS phase. We conclude that fast outflows, responsible for the BAL features, can be more easily present among objects going through a restarting or just-started radio phase, where radiation pressure can substantially contribute to their acceleration.}

\maketitle
\section{Introduction}
Fast outflows are observed in a variety of accreting astrophysical objects, and at different scales, from protostars to Active Galactic Nuclei (AGN). In the latter case, they can have a feedback effect on the host galaxy, as underlined by recent studies (e.g. Feruglio et al. 2010; Wang et al. 2010; Sturm et al. 2011). They can be detected 
in different bands: from the UV/optical, as Broad Absorption Lines Quasars (BAL QSOs, Hewett \& Foltz 2003), to the X-ray band as Ultra Fast Outflows (UFOs) AGNs (Pounds et al. 2003; Tombesi et al. 2014; Fukumura et al. 2015). BAL QSOs are about ~20\% of the total QSO population, and are identified as objects presenting broad absorption (several thousands km/s) in the UV/Optical band, produced by outflows that can reach relativistic velocities (up to 0.2c). In order to explain the observed fraction of such sources some authors proposed a youth scenario (Briggs et al. 1984; Sanders 2002; Farrah et al. 2007), in which BAL QSOs are young objects still expelling a dust cocoon, that can be responsible of the absorption. An alternative scenario proposed by Elvis 2000, explains the phenomenon as an orientation effect, suggesting that the outflow, responsible for the BAL features should be common in QSOs, but visible only when it intercepts the line of sight of the observer. 

Radio-Loud (RL) BAL QSOs are even more rare, in fact objects with $log R^{*}>2$ (radio-loudness, $R^{*}=S_{5 GHz}/S_{2500 \AA}$, Stocke et al. 1992) are only $\sim$25\% of the total BAL QSOs population (Becker et al. 2001). Previous studies of this subclass of sources have inferred different possible orientations for the radio jet, and different morphologies and ages for the radio emission (Montenegro-Montes et al. 2008; DiPompeo et al. 2011; Bruni et al. 2012, 2013), not clearly favouring either of the two scenarios. Also, the Compact Steep Spectrum (CSS) and GigaHertz Peaked Spectrum (GPS) source fraction, indicative of a possible young age for the QSO, was found to be comparable among BAL and non-BAL objects (Bruni et al. 2012).

\section{The observational campaign}
In 2009, we selected a sample of 25 RL BAL QSOs and a comparison sample of 34 RL non-BAL QSOs, with redshift 1.7$<z<$4.7, later presented in Bruni et al. 2012. We conducted an extensive observational campaign, involving different techniques and instruments in the cm-band (GMRT, VLA, Effelsberg-100m single dish, EVN and VLBA continental interferometers), mm-band (IRAM-30m, APEX), and infrared/optical bands (TNG, WHT) to explore the nature of these peculiar objects, and test the proposed scenarios. Thanks to our first radio Spectral Energy Distribution (SED) and polarisation study (Bruni et al. 2012) we could identify a $\sim$30\% of GPS, young, sources in our BAL sample. Some sources ($\sim$12\%) presented an additional, older, radio component below 1.4 GHz. For this frequency range (70 MHz - 408 MHz), we collected ancillary data from surveys. Another $\sim$25\% presented a single component, with no visible peak in our own or the literature data.

In 2010, we started a second session of observations to complement the data previously collected in the cm-band (between 1.4 GHz and 43 GHz) with observations at lower 
frequencies (235 MHz and 610 MHz) using the GMRT interferometer, and in the mm-band (250 GHz, 345 GHz, and 850 GHz) with bolometers at the IRAM-30m (MAMBO2) and APEX (LABOCA, SABOCA) single dishes. Our purpose was to test the presence of extended, relic emission in BAL QSOs already showing emission at low-frequencies from the literature - that would indicate an old radiative age (10$^7$-10$^8$ years, Konar et al. 2006, 2013) for these objects - and probe the possible dust emission in the mm-band - that would suggest the presence of a dust cocoon, as expected in the young scenario.

\section{Results}

\subsection{The low-frequency regime}
\begin{figure}[]
\includegraphics[width=8cm]{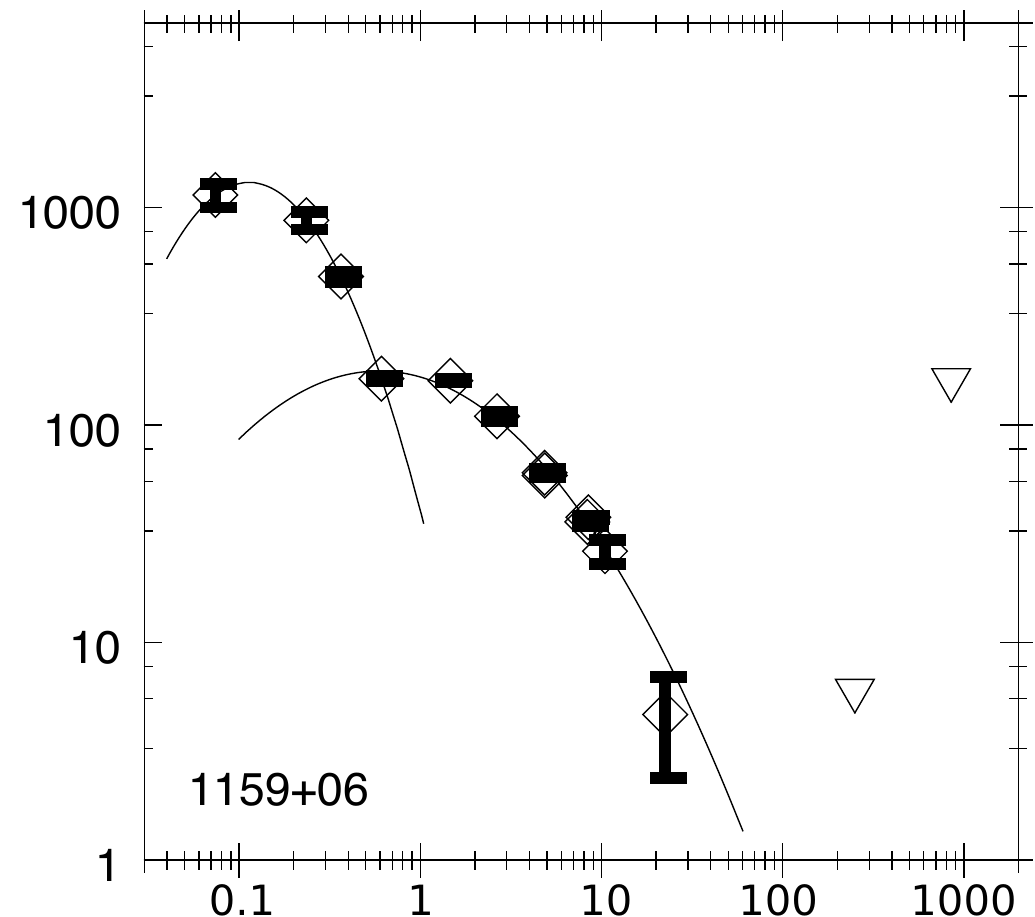}
\caption{SED (mJy \emph{vs} GHz) of one of the RL BAL QSOs presenting a CSS plus a GPS component (Bruni et al. 2015). Diamonds are flux densities collected during our campaign or from surveys, triangles are upper limits from IRAM-30m (250 GHz) and APEX (850 GHz) telescopes.}
\label{double}
\end{figure}

All the five objects ($\sim$25\% of the sample) observed at the GMRT present an extended emission, with a projected linear size of several tens of kpc. This emission is clearly separated from the one identified in the GHz range, unresolved in the previous observations exploring a scale of a few kpc. Moreover, thanks to these data, three sources turned out to have two-component emission: one peaking in the GHz range, and another one in the MHz range (see Fig. \ref{double}). This could indicate a restarting activity, with a new component arising from the core. Other sources from the sample already showed hints of a CSS component peaking in the MHz range, but not all of these could be observed at the GMRT.

\subsection{Dust abundance}
Bolometer observations in the mm-band were conducted to detect dust emission at frequencies from 250 GHz upwards. Thanks to the previously reconstructed synchrotron emission, we aimed at detect the grey-body emission in excess of the synchrotron tail. 
%
%
Considering the recent results from the \emph{Herschel}-ATLAS project about dust emission in RL and RQ QSOs (Kalfountzou et al. 2014), a mean flux density in the range 10-20 mJy would be expected between 250 and 850 GHz for our sources.
We observed about half of the sample with the IRAM-30m single dish, while only equatorial sources could be observed with APEX. Only one source shows a flux density, at 250 GHz, clearly in excess of the synchrotron one, corresponding to 6\% of the observed sources at this frequency. APEX observations resulted in upper limits of $\sim$100 mJy. As a comparison, Omont et al. 2003 presented a study of dust emission in Radio-Quiet (RQ) QSOs, conducted at 250 GHz with the MAMBO bolometer at IRAM-30m. For these observations, RMS values ($\sim$1 mJy) and redshift range (1.8$<z<$2.8) were similar to ours. They found that 26\% of their targets shows dust emission, a fraction four times greater than found in our sample.

\section{Discussion}

Our previous works found mixed properties for BAL QSOs, resulting in a range of possible orientations, different radio ages, and synchrotron polarisation fraction similar to that found in normal QSOs. Also a deeper look at the pc scale, through VLBI techniques, highlights different morphologies (double, core-jet, unresolved; Bruni et al. 2013). Projected linear sizes can range from a few pc to hundreds of kpc, sometimes showing a misalignment between the inner core-jet component and the more extended one.
The radio phase itself does not seem to correlate with substantial differences in the AGN geometry and accretion, resulting in similar Eddington ratios, broad line region radius, and central BH mass for RL and RQ BAL QSOs (Bruni et al. 2014). 

The latest results from our m- and mm-band campaign (Bruni et al. 2015), present some new highlights on the dust abundance, found to be even lower than in normal QSOs. This could suggest a lower star formation rate, possibly inhibited as a feedback effect of the relativistic BAL-producing outflow. This should be confirmed by further observations. The fact that the five objects observed in the MHz range with the GMRT present an extended component, confirm that RL BAL QSOs can have previous radio component coexistent with a younger one peaking in the GHz range. Considering the previously found fraction of GPS in our sample, it turns out that about 70\% of our sample is in a GPS or GPS+CSS phase. This could suggest that fast outflows can be more easily present in object experiencing a transition phase, where a new plasma is being expelled from the core, or during the ignition of the radio phase itself. A favourable condition for the acceleration of the outflow could be the particularly high radiation pressure found during these events.

\section{Future plans}

\subsection{The high-redshift campaign}
Allen et al. 2011 found a dependence with redshift of the BAL fraction, the latter increasing with redshift. This suggests a cosmological evolution of this kind of objects. We have conducted an observational campaign on a further sample of high redshift ($z>3.6$) RL BAL QSOs, to study their radio properties (Tuccillo et al. 2015). Observations with both JVLA and Effelsberg-100m have been performed. If a majority of GPS/CSS+GPS sources will be found, it would support the hypothesis that BAL QSOs are common in newborn or restarting radio sources.

\subsection{Optical/X-ray properties and possible connection with UFOs}

In 2012 we also collected intermediate resolution spectra of our sample at the WHT, in order to perform a detailed study of the outflow properties. We recently proposed \emph{Chandra} snapshots of the most X-ray luminous objects (5 RL and 5 RQ), to determine the X-ray brightness between 2 and 10 KeV. This could help in disentangling the contribution of the radio phase in the overall BAL phenomenon, and test the $\rm{C}_{\rm{IV}}$/X-brightness relation using the WHT spectra.

\section{Conclusions}

From our previous and latest observational campaigns, we can conclude that:

\begin{itemize}

\item RL BAL QSOs can present a variety of morphologies, radio ages, and orientation. From our previous studies in the cm-band, they did not seem particularly young objects, the overall GPS fraction being $\sim$30\% of the total. None of the scenarios presented in the literature seemed to be favoured by these results. However, additional observations in the low frequency regime ($<$1.4 GHz), show that they can have extended components, peaking at hundreds of MHz (CSS) or lower frequencies. In some cases a GPS component coexists with the previous one, suggesting a turbulent phase in which a new, young, component is being expelled from the core. About 70\% of our sample is in a GPS or CSS+GPS phase.

\item Observations in the mm-band show that BAL QSOs have less dust emission, and thus probably less dust mass, with respect to normal QSOs. This could be explained by a feedback effect of the BAL-producing outflow on the host galaxy, resulting in a lower star formation rate (to be proved by further observations).

\item Results from our work seem to indicate that outflows present in BAL QSOs can be more easily launched in objects going through a new radio phase, or during first ignition of the radio phase itself, where favourable high radiation pressure is present.

\end{itemize}

\acknowledgements
Part of this work was supported by a grant of the Italian Programme for
Research of National Interest (PRIN No. 18/2007, PI: K.-H. Mack)
The authors acknowledge financial support from the Spanish Ministerio de Ciencia e Innovaci\'on under project 
AYA2008-06311-C02-02.
%
%
%
We thank the staff of the GMRT that made these observations possible. GMRT is run by the National Centre for Radio Astrophysics of the Tata Institute of Fundamental Research.
This publication is based on data acquired with the Atacama Pathfinder Experiment (APEX). APEX is a collaboration between the Max-Planck-Institut f\"ur Radioastronomie, the European Southern Observatory, and the Onsala Space Observatory.
This work is partly based on observations carried out with the IRAM-30m Telescope. IRAM is supported by INSU/CNRS (France), MPG (Germany) and IGN (Spain).
%



\end{document}